# Automated Dynamic Concurrency Analysis for Go

Saeed Taheri

University of Utah
Utah, USA
staheri@cs.utah.edu

Ganesh Gopalakrishnan

University of Utah
Utah, USA
ganesh@cs.utah.edu

The concurrency features of the Go language have proven versatile in the development of a number of concurrency systems. However, correctness methods to address challenges in Go concurrency debugging have not received much attention. In this work, we present an automatic dynamic tracing mechanism that efficiently captures and helps analyze the whole-program concurrency model. Using an enhancement to the built-in tracer package of Go and a framework that collects dynamic traces from application execution, we enable thorough post-mortem analysis for concurrency debugging. Preliminary results about the effectiveness and scalability (up to more than 2K goroutines) of our proposed dynamic tracing for concurrent debugging are presented. We discuss the future direction for exploiting dynamic tracing towards accelerating concurrent bug exposure.

## 1 Introduction

Go [4] is a statically typed language initially developed by Google and at present widely used by many. It employs channel-based Hoare's Communicating Sequential Processes (CSP) [23] semantics in its core and provides a productivity-enhancing environment for concurrent programming. The concurrent model in Go centers around 1) *goroutines* as light-weight user-level threads (processing units), 2) *channels* for explicit messaging to synchronize and share memory through communication, and 3) a *scheduler* that orchestrates goroutine interactions while shielding the user from many low-level aspects of the runtime. This design facilitates the construction of data flow models that efficiently utilize multiple CPU cores. Because of the simple yet powerful concurrency model, many real production software systems take advantage of Go, including container software systems such as Docker [29], Kubernetes [9], key-value databases [5], and web server libraries [8].

In traditional shared-memory concurrent languages such as Java/C/C++, threads interact with each other via shared memory. Processes in CSP-based languages such as Erlang communicate through mailbox (asynchronous) message passing. Go brings all these features together into one language and encourages developers to *share memory through communication* for safe and straightforward concurrency and parallelism. The visibility guarantee of memory writes is specified in the memory model [10] under synchronization constraints (*happens-before* partial order [26]). The language is equipped with a rich vocabulary of *serialization* features to facilitate the memory model constraints; they include synchronous and asynchronous communication (either unbuffered or buffered channels), memory protection, and barriers for efficient synchronization. This rich mixture of features has, unfortunately, greatly exacerbated the complexity of Go debugging. In fact, the popularity of Go has outpaced its debugging support [7, 40, 44]. There are some encouraging developments in support of debugging, such as a data race checker [42] that has now become a standard feature of Go, and has helped catch many a bug. However, the support for "traditional concurrency debugging" such as detecting atomicity violations and Go-specific bug-hunting support for Go idioms (e.g., misuse of channels and locks) remain insufficiently addressed.

In this work, we present the initial steps that we took towards addressing this lack. Since a bug might occur at various levels of abstraction, dynamic tracing provides a practical and uniform way to track multiple facets of the program during execution (as we have shown in our prior work [36]). Also, unlike assertion-based tools [28, 43], a dynamic tool is more automated, not requiring user expertise. We developed a facility that automatically gathers *execution concurrency traces* (i.e. sequences of events) during the execution of Go applications with minimal instrumentation. By enhancing the Go built-in tracing mechanism with *concurrency usage* events, we enrich original *execution traces* so that they accurately reflect the dynamic concurrency behavior of applications. Upon Go programs' termination when tracing is enabled, traces are flushed and structurally stored in relational tables of an SQL database, enabling multi-aspect program analysis in offline.

With the help of this novel *automated dynamic tracing* mechanism, our testing framework under development *accelerates* bug exposure by manipulating the native scheduler around *critical points* in the code—combination of constructs that heighten the propensity for bug-introduction (more in section 4). Our work in progress aims to enhance interleaving coverage around such critical points, thus increasing the likelihood of unearthing bugs. While these ideas have been developed and proved to be effective in other contexts [12, 17, 30], our contribution is to show these ideas in the context of a modern language with growing industry-side adoption.

To summarize, here are our contributions:

- We take the tracing mechanism embedded in the standard Go that captures *execution trace* (ET) and enhance it with a set of concurrency primitive usage events to obtain *execution concurrency trace* (ECT). While the primary usage of ET is performance analysis, ECT provides an accurate and comprehensive model of concurrent execution, enabling automated analysis of logical behavior and concurrent bug detection.
- We introduce a framework that automatically instruments the target program, collects ECT, and structurally stores them in a database. Through querying the database, several visualizations and reports are accessible.
- We propose an approach to identify the points (i.e. source line number) in the target program in which a random noise might drastically change the program's dynamic behavior. We analyze ECTs to identify such points and measure schedule space coverage per execution.

The rest of this paper is as follows: Section 2 discusses correctness problems and approaches in Go. Section 3 describes the enhancement we made to the tracer package. Section 4 discusses the advantages of dynamic tracing for concurrent debugging and draws the ongoing and future direction of the current work. At last, section 5 summarizes and concludes.

## 2   Correctness in Go

Go introduces a new concurrency model, mixing shared-memory and message-passing concepts with an ad-hoc scheduler:

- **Goroutines** are functions that execute concurrently on logical processors having their own stack.
- **Channels** are typed conduits through which goroutines communicate. Channels are unbuffered by default, providing synchronous (rendezvous) messaging between goroutines.
- **Synchronization** features such as *mutex*, *waitGroup*, *conditional variables*, *select*, and *context* are included in the language to provide more and flexible synchronization, data access serialization, memory protection, and error handling.
- **Scheduler** maintains goroutines in FIFO queues and binds them on OS threads to execute on processing cores.

Listing 1 shows a simplified version of a reported bug in Docker [1]. An instance of the Container

**Listing 1** Simplified version of bug moby28462

```
1   package main
2   import "sync"
3
4   type Container struct{
5       sync.Mutex
6       stop chan struct{}
7   }
8
9   func main() {
10      container := &Container{
11          stop:make(chan struct{})}
12      go  Monitor(container)
13      go  StatusChange(container)
14  }
```

```
15  func Monitor(cnt *Container){
16      for{
17          select{
18          case <- cnt.stop:
19              return
20          default:
21              cnt.Lock()
22              cnt.Unlock()
23  }}}
24  func StatusChange(cnt *Container){
25      cnt.Lock()
26      defer cnt.Unlock()
27      cnt.stop <- struct{}{}
28  }
```

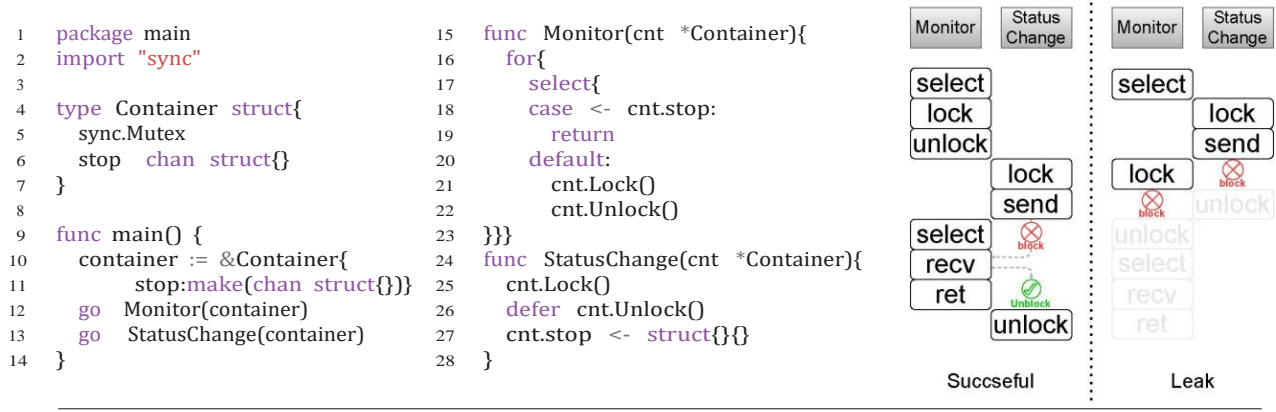

type (lines 4-7) is created in the main function (lines 10-11). In line 12, a goroutine is spawned to execute function Monitor that continuously checks the container status and returns once it receives from the container's channel (lines 18-19). The default case of the select statement (line 20) allows Monitor to continue monitoring without getting blocked on the channel receive (line 18). Concurrent to the main and Monitor goroutines, another goroutine is created in line 13 to execute function StatusChange which changes the status of the container by sending to the container's channel. The container's lock is released after the send action completes and function returns (defer statement in line 26).

Native execution of this program successfully terminates without issuing any error or warning. Based on the Go specification and memory model, there is no constraint on the goroutines spawned from the main function to join back before the main goroutine[1] terminates. A deadlock detector within the runtime periodically checks that the scheduler queues of all *runnable* goroutines never become empty until the main goroutine terminates. In other words, the runtime throws a deadlock exception when the main goroutine is blocked, and no other goroutine is in the queue to execute (i.e. *global deadlock*). Since there is no blocking instruction in the main goroutine in listing 1, the program terminates successfully regardless of other goroutines' statuses. However, this program suffers from a common bug in concurrent Go where one or more goroutines *leak* (i.e. *partial deadlock*) from the execution (i.e. never reach their end states).

Due to the non-determinism introduced by the runtime scheduler and application-level random features like select, many interleavings are feasible for concurrent execution of simple programs such as listing 1. The right side of the listing displays a successful and a *leaky* interleaving of the program. In the leaky scenario, first, the Monitor goroutine executes the select statement and, based on the available cases, picks the default case to execute. Right before the execution of mutex lock (line 21), the scheduler context-switches and the StatusChange goroutine starts its execution through which it holds the lock and blocks on sending to the channel (line 27) since there is no receiver on that channel. Upon blocking on send, the scheduler transfers back the control to the Monitor goroutine that tends to acquire the mutex, but because the mutex is already held by StatusChange, the Monitor goroutine also blocks. The circular wait between the container mutex and channel prevents both spawned goroutines from reaching their end states and leaves the program in an unnoticed deadlock situation. Widely used deadlock detectors such as Goodlock [22] are not applicable in Go since causes of Go deadlocks are resources (e.g. locking a locked mutex) or communication (e.g. sending on a full channel), or a combination of them (e.g. leaky interleaving of listing 1). In addition, due to the light-weight nature of goroutines, it is not uncommon to spawn thousands of goroutines in production software systems. Hence, novel and scalable techniques are needed to enable realistic modeling of program behavior during execution.

---

[1] In the remainder of the paper, we use *main function* and *main goroutine* interchangeably.

Decades of research effort have been dedicated to the logical and performance correctness of concurrent and parallel programs. For CSP-based concurrent languages like Go, static (source-level) analysis methods [27, 28, 31] tend to assure bug freedom and verify safety properties through abstractions like session types and choreography synthesis. Dynamic (runtime-level) analysis approaches [14, 35, 42] rely on code instrumentation and program rewrites to obtain and analyze an *execution model*. Although these methods show effectiveness in analyzing specific aspects of correctness analysis, they usually do not scale for real-world Go applications with thousands of goroutines and LOC [13].

It is crucial for debuggers and software analysis tools to construct their abstract models as close as possible to the actual program execution context. For multi-threaded Java, effective dynamic methods like ConTest [16], Goodlock [22] and CalFuzzer [25] maintain a model built from *synchronization constructs* of the program, as the main ingredient of dynamic concurrency model. Our investigations state that such comprehensive dynamic data collection mechanisms to abstract reliable concurrency models for Go applications do not exist. Profilers [32] give up some accuracy by approximating the dynamic behavior through aggregated samples from counters, while distributed (decentralized) tracing systems [33] gather logs and information (e.g. HTTP request latency) through source instrumentation and an underlying network.

We found that Go execution tracer [19] gives detailed information about the *performance* behavior during execution. Its tracing capability is compiled into all programs always through the runtime and is enabled on demand – when tracing is disabled, it has minimal runtime overhead [41]. An execution model from the interactions of processors, OS threads, goroutines, the scheduler, and the garbage collection mechanism is constructible from the trace to identify poor parallelization and resource contention. By enriching the tracer package with concurrency events, we place the missing pieces for human debuggers and analysis tools to automatically obtain comprehensive models from the dynamic concurrency behavior of programs with minimal overhead.

## 3 Automatic Dynamic Tracing

According to [40, 44], the primary cause of most real-world bugs is the misuse of concurrency features like channels, mutexes, and waitGroups. However, the standard tracer package does not capture concurrency primitive usage events as it aims more on performance bugs. Although the event vocabulary is rich enough to model comprehensive goroutine latency and blocking behavior accurately, the tracer package cannot answer questions such as do the synchronization constructs work as expected? In the leaky interleaving of listing 1, which goroutine is holding the lock that Monitor tends to acquire? What are the resources that contribute to this circular wait? In the successful interleaving of listing 1, is there any *potential flaw* that did not manifest?

| Category | Description |
|----------|-------------|
| Process | Indication of process/thread start and stop |
| GC/Mem | Garbage collection and memory operation events |
| Goroutine | Goroutines events: create, block, start, stop, end, etc. |
| Syscall | Interactions with system calls |
| Users | User annotated regions and tasks (for bounded tracing) |
| Misc | System related events like futile wakeup or timers |

Table 1: Event categories by the Go execution tracer

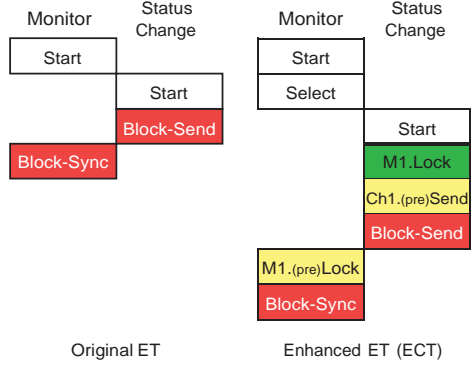

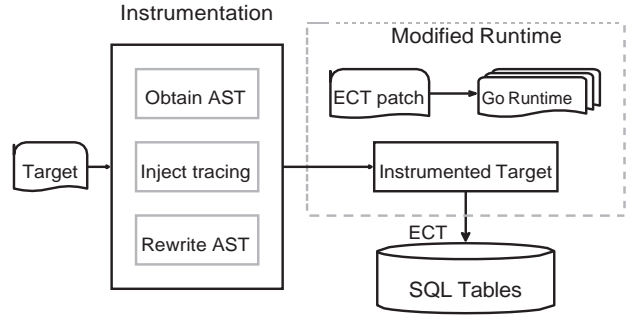

Figure 1: Reflection of leaky situation of listing 1 in ET vs. ECT

Figure 2: Framework Overview

As our previous works stated [36, 37], dynamic tracing provides a practical and uniform way to answer the above debugging questions and track multiple program facets during execution. In the context of Go, we propose a framework (figure 2) that automatically captures dynamic concurrency behavior exploiting an enhancement to the tracer package. As a result, we chose the tracer package to enhance because it 1) is already compiled into the runtime, 2) adds minimal overhead, and 3) only lacks some pieces allowing the construction of an accurate concurrency model.

## 3.1 Execution Concurrency Tracing (ECT)

When enabled, the execution tracer package from the standard Go captures and compresses an *execution trace* (ET). Upon program termination, the ET is flushed to an IO buffer. The decompressed ET is a totally ordered *sequence* of events in which the order is approximated through a central clock with nanosecond precision. ET also contains the call-stack for each event enabling a direct mapping of events to source-line numbers. The alphabet of ET is total of 49 events [2], categorized and summarized in table 1. The left-side diagram in figure 1 is an *execution model* constructed from the ET of the leaky interleaving in listing 1.

Although models obtained from ET reflect the dynamic blocking behavior of programs, ET is not able to provide insight into the *concurrent state* of the program (i.e. concurrency actions that each goroutine has performed or tends to perform) at any given execution step. Through a one-time patch, we enrich the original tracer package to emit *execution concurrency traces* (ECT) by extending the vocabulary of ET to capture concurrency usage events:

- **Channel:** For each channel operation (make, send, receive, close), ECT includes an event with a unique id assigned to each channel.
- **Mutex, WaitGroup & Conditional Variables:** Similar to channels, we assign a unique id to each concurrency object and emit an event for each of their operations (lock, unlock, add, wait, signal, broadcast).
- **Select & Schedule:** The scheduler and the *select* structure introduce non-determinism to the execution. We keep track of the decisions made by the scheduler and select statements to obtain an accurate decision path during execution.

Blocking concurrency operations such as channels *sends/recvs*, mutex *locks*, waitGroup/conditional variable *wait* and *select* (when none of the cases are available) prevents the executor goroutine from making progress. For each blocking operation, ECT captures a pair of pre-operation and post-operation events (yellow and green boxes respectively). Hence, ECT enables concurrency state modeling at any given step of execution. In section 4 we discuss the effectiveness and feasibilty of such tracing mechanism towards debugging concurrent Go.

## 3.2   Automatic ECT Collection

As shown in figure 2, we obtain the source abstract syntax tree (AST) using the Go AST package [20] to inject *tracing handlers* into any given target program. Tracing handlers are function invocations to enable/disable tracing and a *watcher* goroutine to monitor the progress of tracing. The watcher flushes out the trace buffer after a pre-defined timeout in case of a global deadlock. We should note that the tracing might perturb the native program or Go runtime interleaving of goroutines. However, it preserves the original semantics of the system such as happens-before partial order between events; thus ECTs remain a reliable source for studying applications concurrency behavior.

The execution of automatically instrumented target program within the modified runtime would result in encoded (compressed) ECT. The decompressed ECTs from real-world applications are often long, with hundreds of entries for each event's stack-frames and arguments. To store ECTs structurally and systematically analyze them in offline space, we insert the decompressed ECTs into SQL relational tables. For each ECT, we store the sequence of events in *Events* table[2], and auxiliary *StackFrames* and *Arguments* tables with *Events.id* as foreign keys. Structured storage of data enables various execution model creation from the program execution through a query-and-replay operation (e.g. vaisualizations such ash figure 1). Next section assesses the effectiveness and feasibility of ECT towards debugging concurrency issues in Go. We are actively working on releasing the full feature tool for wide usage in the community. A prototype of the framework is available in [3].

# 4   Discussion

In this section, we discuss the idea of ECT by enumerating some of its significant advantages towards debugging and quantify its collection cost for large-scale programs. The section ends with some introduction to the novel usage of ECTs.

## 4.1   Debugging via ECT

ECT enables the production of precise and comprehensive concurrency execution models. For example, the right side of figure 1 visualizes an execution model constructed from the captured ECT of leaky execution in listing 1. The ECT clearly and precisely reflects the leaky execution where the Monitor goroutine is blocked on locking the mutex $M1$ while StatusChanged already holds $M1$. Other visualizations such as resource wait-for graph, happened-before replay of events through Shiviz [11] and execution visualizations like figure 1 are constructible from ECTs. Human debuggers can use such models to review program execution and compare their expectations with actual behavior. Moreover, online or offline verifiers can take ECT as input and construct logical models to verify the program execution.

---

[2]A description of the relational tables is available in [3].

**Listing 2** Concurrent Prime-seive (bounded by N)

```
1   package main                          11   func main() {
2   import "os"                           12     N := int(os.Args[1])
3   func Generate(ch chan int) {          13     ch := make(chan int)
4     for i := 2; ; i++ { ch <- i }       14     go Generate(ch)
5   }                                      15     for i := 0; i < N ; i++ {
6   func Filter(in, out chan int, prime int){  16       prime := <-ch
7     for {                               17       ch1 := make(chan int)
8       i := <-in                         18       go Filter(ch, ch1, prime)
9       if i%prime != 0 { out <- i }      19       ch = ch1
10    }                                   20     }
11  }                                     21   }
```

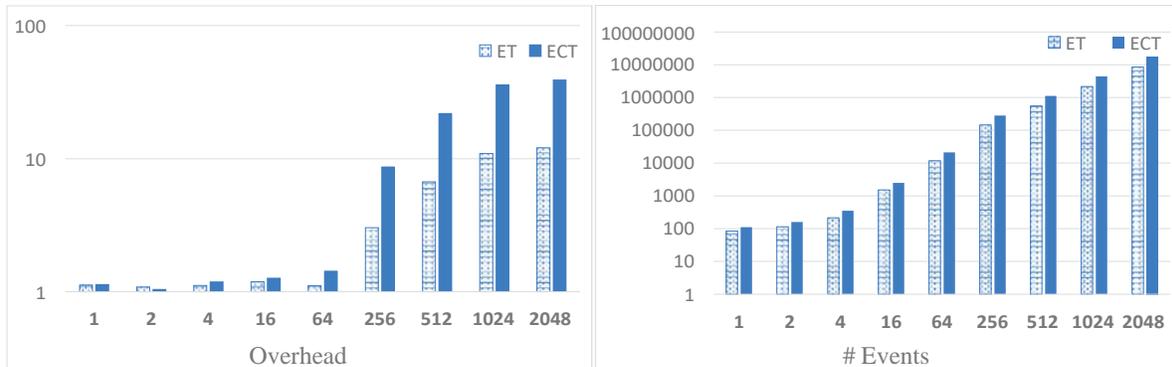

Figure 3: Evaluation of ET vs. ECT (left: overhead added to the native, middle: number of events

While tracing is enabled during the execution of Go programs, the tracer records events for all goroutines, including the Go runtime system, tracing system, and *application-level* goroutines. We say a goroutine is an application-level goroutine if it is the main goroutine (that executes the main function) or it has all of the following conditions: 1) its ancestor is the main goroutine, 2) it is not a Go runtime system goroutine, and 3) it is not a tracer goroutine. Identifying the application-level goroutines is done by multi-dimensional queries to the database to reason about the creation site of each goroutine. Such distinguishment between goroutines is essential to define the boundaries of the application and the underlying system. For example, one can infer that "an application-level goroutine is leaked" during execution if the final captured event of its ECT is something other than GoEnd.

### 4.2 Overhead

The program in listing 2 is an implementation of concurrent "prime-sieve" that we borrowed from [27] to output first N prime numbers. The Generate function continously send incremental sequence of integers starting from 2 to its sender channel ch (line 4). Filter receives values from the receiver channel in (line 8) and sends the values that are not divisible by the previously found prime number to out (line 9). The main function bootstraps the sequence generator and outputs the first N prime numbers through a for loop (lines 15-20). Basically, main creates a pipeline of channels and goroutines where each spawned goroutine (a concurrent instance of *Filter*) is responsible for sieving the receiving numbers divisible by the previously found prime number.

The concurrent behavior of this simple program is quite complex and challenging to reason about its correctness. It creates channels and spawns goroutines dynamically during execution relative to the input *N*. The state space of such an intense orchestration of goroutines and channels explodes quickly with the increase of *N*. We performed a set of experiments to evaluate the feasibility and scalability of our tracing idea by enabling ET and ECT collection over the execution of concurrent prime-sieve.

| INPUT | # | # | Overhead | | # Events | | Trace Size(KB) | |
|-------|-----|------|------|-------|-----------|------------|-----------|-------------|
| N | GRTN | CHNL | ET | ECT | ET | ECT | ET | ECT |
| 1 | 10 | 4 | 1.12 | 1.13 | 85 | 112 | 2.68 | 3.61 |
| 2 | 11 | 5 | 1.08 | 1.04 | 114 | 161 | 2.83 | 4.02 |
| 4 | 13 | 7 | 1.11 | 1.18 | 214 | 355 | 3.21 | 5.08 |
| 16 | 25 | 19 | 1.19 | 1.26 | 1,503 | 2,516 | 7.89 | 18.20 |
| 64 | 73 | 67 | 1.11 | 1.43 | 11,750 | 21,336 | 45.55 | 140.91 |
| 256 | 265 | 259 | 3.02 | 8.65 | 145,036 | 283,339 | 559.07 | 2,031.50 |
| 512 | 521 | 515 | 6.69 | 21.78 | 552,374 | 1,116,021 | 2,163.63 | 8,327.83 |
| 1024 | 1033 | 1027 | 10.90 | 35.84 | 2,157,613 | 4,420,998 | 8,442.12 | 33,745.92 |
| 2048 | 2057 | 2051 | 12.04 | 39.00 | 8,521,690 | 17,594,298 | 33,313.36 | 140,862.68 |

Table 2: Breakdown of scalability measurements. Each row is the average number after 100 executions. Columns descriptions (left to right): **INPUT N**: number of loop iterations in concurrent prime-sieve, #**GRTN**: number of dynamically spawned goroutines, #**CHNL**: number of dynamically created channels, **Overhead:** the ratio of run-time when tracing is enabled over native execution, #**Events**: length of the captured sequence of events, **Trace Size:** the size of encoded trace files on disk (KB)

The experiments are performed on a MacBookPro with 2.5 GHz Intel Core i5 processor and 16 GB 1600 MHz DDR3 memory. The behavior of tracing (overhead, length of captured sequences and trace size on disk) are shown in table 2 and figure 3. There is a steep increase in the overhead added to the native program when N jumps from 64 to 256. This overhead jump is the exponential number of context-switches during execution, which is required to distribute computation power among many simultaneous goroutines equally. For each context-switch, the original ET (and consequently ECT) contains a set of events describing the preemption/blocking/unblocking and (re)start of goroutines. Hence, when the number of spawned goroutines increases to a number much higher than available processing cores, the load on the tracer package increases drastically to emit all context-switch events.

## 4.3   Ongoing Work

While the common bug patterns such as listing 1 are well known, their occurrence within the context of large codebases and mixed with multiple event-types might get overlooked or lost. Assuming the scheduler's non-determinism is the only cause of concurrency bugs (and not the program input), even after comprehensive unit testing, the Go scheduler might never execute the leaky interleaving to reveal the flaw. We are implementing a testing framework that injects random delays around source locations that increases the probability of executing an untested interleaving. The literature [12, 17, 30, 38] and our preliminary experiments show that systematic manipulation of schedulers is an effective approach that accelerates the exposure of hidden concurrent bugs.

In general, a context-switch or preemption might happen at *any* point of the program. However, only context-switches that occur around concurrency primitive usages change the scheduler and blocking behavior of the program. Adopting ideas from existing concurrency testing tools, we want to navigate the scheduler towards executing likely-erroneous interleaving systematically. As explained in section 3, ECT enables accurate construction of the concurrency model of the program by capturing an event per concurrency primitive usage. By extracting source locations of concurrency usages from ECTs, we can identify the points in the target program that a random delay might perturb the program's dynamic behavior. We refer to those points as "critical points" of the program. By inserting *random* and *bounded* scheduler perturbation calls around these critical points (using the AST package), we enforce the program to context-switch and take a potentially untested execution path. Moreover, Concurrency coverage metrics such as *sync-pair* [24], *blocking-blocked* [15], and *blocked-pair* [39] that focus on synchronization bugs such as deadlocks are measurable using ECTs. We are implementing their algorithms to quantify the quality of each execution (e.g., measuring coverage of execution schedule) under the scheduler perturbation mechanism.

While the idea of schedule alteration is nothing new, making it practical in a production language such as Go brings its challenges. Ours is the first (to the best of our knowledge) practical approach (with implementation underway) for Go that automatically identifies potential pitfalls in the program and exercises the scheduler around them, relying on extending a provided tracing mechanism. The idea of critical points is motivated by the fact that unlike algorithms such as dynamic partial order reduction [18] which go by a fixed notion of dependency, critical points are those that are points of execution where a context-switch can cause an interesting alternate schedule. In a sense, at these points, a context-switch can be forced by a yield action whose probability can be chosen based on various criteria [12, 17, 30].

Here are some advantages of the proposed idea. First, we only manipulate the scheduler around points that can fall into a buggy interleaving. It maximizes the probability of executing buggy scenarios (exposing bugs) and decreases the interleaving state by ignoring unpopular context-switches. Second, we can improve the captured heuristic by automatically re-executing programs under a manipulated scheduler and increase the accuracy of critical points' identification. Third, due to the tracer package's feasible overhead and shortcomings of current debugging tools, this framework is promising and practical. We can add to these the capability of covering and capturing dynamic goroutine and channel creations, which are the limitations of the existing tools.

## 5   Summary and Conclusion

Stat-of-the-art concurrency analysis tools for Go are shorthanded in supporting large production programs with dynamic goroutine and channel creation. The lack of a dynamic analysis infrastructure for Go concurrent programs motivates us to extend the standard tracer of the language and construct execution models of concurrent program behavior. Such models allow debuggers and analysis tools to view concurrency behavior during program execution and use that knowledge for various correctness inferences. For instance, our primary experiments show that we can detect bugs and their root causes in all of GoKer [44] blocking benchmarks by analyzing their ECTs. A prototype of this tracing framework is available for the community for practical use and contribution.

Due to the non-deterministic nature of concurrent programs, concurrency bugs might get overlooked in the testing phase as the tested execution paths might never trigger the bug. Dynamic tracing of concurrent programs helps identify critical context-switch points during execution such that altering interleavings around those points accelerates bug exposure. Currently, we are working on a testing framework to perturb the native scheduler around critical points (extracted from ECT) for systematic interleaving exploration. Our preliminary results are promising as we were able to accelerate bug triggering scenarios for many hard-to-trigger bugs bugs in the benchmark suite [44].